\documentclass[showpacs,12pt]{revtex4}

\usepackage{graphicx}
\usepackage{amsmath}

\begin{document}

\title{Braided nodal lines in wave superpositions}

\author{M R Dennis}

\affiliation{H H Wills Physics Laboratory, Tyndall Avenue, Bristol BS8 1TL, UK}

\begin{abstract}
Nodal lines (phase singularities, optical vortices) are the generic interference fringes of complex scalar waves. Here, an exact complex solution of the time independent wave equation (Helmholtz equation) is considered, possessing nodal lines which are braided in the form of a borromean, or pig-tail braid. The braid field is a superposition of  counterpropagating, counterrotating, non-coaxial third order Bessel beams, and a plane wave whose propagation is perpendicular to that of the beams. The construction is structurally stable, and can be generalized to a limited class of other braids.
\end{abstract}

\maketitle

In three dimensional fields of interfering complex scalar waves, perfect destructive interference, that is, where the total wave amplitude is zero, generically occurs along lines: nodal lines.
They are also called phase singularities, since the phase is undefined when the amplitude is zero, or wave dislocations \cite{nb:34,nye:natural}.
Nodal lines in waves are much studied, particularly in solutions of the time independent (Helmholtz) wave equation, in which they are stationary features of the interference pattern. 
They appear in a variety of physical contexts, including quantized vortex lines in solutions of the Schr{\"o}dinger equation in quantum mechanics \cite{dirac:singularities},  and optical vortices in scalar components of monochromatic electromagnetic fields \cite{nye:natural, sv:singular}, where their presence in optical beams often indicates the presence of orbital angular momentum \cite{absw:laguerre}.
Generally, in interfering spatial fields, there are very many nodal lines, tangled in a complicated way \cite{bd:321}, reminiscent of the tangling of quantized vortices in media such as turbulent superfluid helium \cite{feynman:application}.

\begin{figure}
\begin{center}
\includegraphics*[width=4cm]{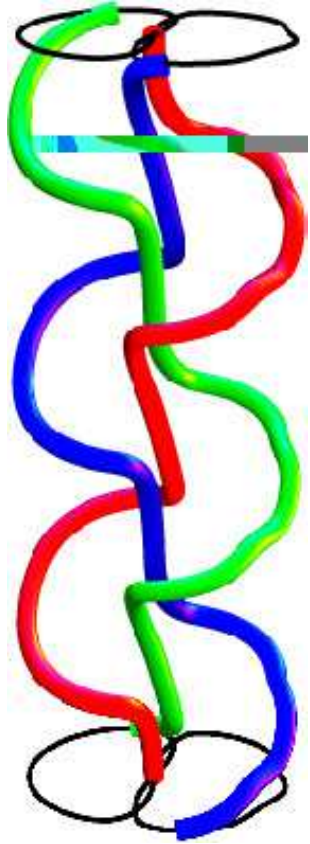}
\end{center}
    \caption{Braided nodal lines in the field described by equations (\ref{eq:braid}), (\ref{eq:pars}). The black curves represent the $x,y$ trajectories of the braid strands, and the structure is plotted to scale. Three $z$-periods are plotted, between $z = 0$ and $6\pi/k_z.$}
    \label{fig:braid}
\end{figure}

Here, I shall demonstrate that a particular complex wave superposition (involving two Bessel beams, defined in equation (\ref{eq:bessel}) below, and a plane wave) contains {\em braided} nodal lines.
As shown in figure \ref{fig:braid}, the field has three phase singularities whose trajectory, as $z$ increases, traces out a figure-8 in the transverse $xy$-plane, forming the well-known borromean braid (also called the pigtail braid, or plait).
This crafting of waves with a braided nodal configuration complements wave fields containing knotted and linked nodal lines in free space \cite{bd:332,bd:333} and in the quantum mechanical electronic states of hydrogen \cite{berry:328}. 
Braiding of nodal lines in light also complements Ref.\ \cite{rhfdm:mutual}, in which two laser beams in a plasma were twisted into a double helix; the topology here is more complicated (the transverse trajectory being a figure-8, rather than a circle), and occurs in linear wave superpositions in free space.
The topology of braided nodal lines is reminiscent of braided vortices in topological (magneto-) hydrodynamics \cite{berger:energy}, and braided vortex trajectories in hamiltonian mechanics \cite{berger:hamiltonian}.

It is easy mathematically to find complex scalar fields which contain various arrangements of nodal lines \cite{freund:trajectories}, as products of separate complex functions for each  nodal line. For each such line, two independent real functions $\xi(\mathbf{r}), \eta(\mathbf{r}),$ are found whose zero contours cross on the desired line; the function $\xi + i \eta$ has a nodal line in this position. For example, the field
\begin{equation}
\psi_{\rm{Liss}} = \prod_{j=0}^{2} (x - \cos(z + 2\pi j/3) + \mathrm{i} (y - \sin 2(z + 2\pi j/3))
   \label{eq:lissbraid}
\end{equation}
has three nodal lines, and as $z$ increases, each executes a simple figure-8 Lissajous trajectory in the $xy$-plane. 
The total field is the product of three complex fields; it has the zeros of each of its separate product terms, and the nodal structure of the total field is braided.

However, such fields are not, in general, physically realisable as waves. That is, the field  (\ref{eq:lissbraid}) does not satisfy the time independent Helmholtz equation
\begin{equation}
   \nabla^2 \psi + k^2 \psi = 0.
   \label{eq:helmholtz}
\end{equation}
Nodal lines are usually studied in solutions of this equation because they are stationary, even if the field is changing periodically in time (as in monochromatic optical fields). 
In the following, the wavenumber $k$ will be set to 1.
The most familiar set of solutions to (\ref{eq:helmholtz}) are plane waves $\exp(\mathrm{i} \mathbf{k}\cdot \mathbf{r})$ where $|\mathbf{k}| = 1;$ these do not themselves have phase singularities, although superpositions of them can.

Associated with phase singularities is a topological number, the topological current (or strength). 
This is an integer, equal to the total phase change around the singularity divided by $2\pi.$ 
The right-hand sense of phase increase around the line endows it with a direction; if the nodal line crosses a plane at a nodal point, the associated topological charge is positive if the sense of the line is out of the plane, negative if into \cite{berry:296}.
A convenient set of solutions of (\ref{eq:helmholtz}), which contain a strength $m$ phase singularity up the $z$-axis are the {\em Bessel beams} \cite{durnin:nondiffracting, dme:diffraction, orss:propagation}
\begin{equation}
   \psi_{m}(k_z,x,y,z) =  J_m \left(\sqrt{1-k_z^2} \sqrt{x^2 + y^2}\right) \left(\frac{x+ \mathrm{i} y}{x-\mathrm{i} y} \right)^{m/2} \exp(\mathrm{i} k_z z).
   \label{eq:bessel}
\end{equation}
The factor $\sqrt{1-k_z^2}$ in the argument of the order $m$ Bessel function $J_m$ is the transverse component of the wavevector (fixed by $k_z,$ since the total wavenumber is 1); the Bessel function has a nodal line along $x = y = 0.$
The ratio of $x+\mathrm{i} y$ and $x-\mathrm{i} y$ to the $m/2$ power in  (\ref{eq:bessel}) gives an order $m$ phase singularity: if $m > 0,$ the direction of the topological current is in the $+z$-direction, if $m < 0,$ it is in the $-z$-direction. 
The beam itself propagates in $\pm z$ according to the sign of $k_z.$ 
An advantage of using these nondiffracting singular beams here (instead of, for example, Laguerre-Gauss beams) is that their structure in the $z$-direction is periodic with period $2\pi/k_z.$

High order singularities, when $|m|>1,$ are not structurally stable. If a Bessel beam (\ref{eq:bessel}) is perturbed by a function which breaks its azimuthal phase pattern, the singularity generically unfolds to $|m|$ unit strength nodes, wound as a helix; the singularities are convected by the twisted phase structure of the beam. 
Unlike hydrodynamic vortices, phase singularity vortices do not repel, but may cross, in a manner equivalent to reconnection of vortices in superfluids \cite{kl:vortex}. The crossings themselves are not stable, and, on perturbation, break in either of the two ways that conserve topological current \cite{bd:333}. 

The knot construction of Ref.\ \cite{bd:332} employed a superposition of $n(n+1)/2$ parallel, coaxial Bessel beams of the same order $m,$ with different $k_z$ and coefficients, in order to make a strength $n$ nodal loop at fixed radius from the beam axis, threaded by the strength $m$ axial nodal line.
On perturbation by a $J_0$ beam, the loop unfolds to give an $(m,n)$ torus knot or link, which winds around the $z$-axis $m$ times as it winds around the original loop position $n$ times.
The braid construction here employs a similar technique, by superposing parallel Bessel beams, in this case counterpropagating and non-coaxial, to get a degenerate configuration, which unfolds to a braid after perturbation by an appropriately chosen plane wave.

The solution $\psi_{\text{braid}}$ of (\ref{eq:helmholtz}), containing the braided nodal lines in figure \ref{fig:braid}, is given by
\begin{equation}
   \psi_{\text{braid}}(x,y,z) = 
   \psi_3(k_z, x + d, y, z) + \psi_3(-k_z, x - d, y, z) \\+ \mathrm{i} \varepsilon \exp(-\mathrm{i} y);
   \label{eq:braid}
\end{equation}
where the parameters $\varepsilon, d$ and $k_z$ have values
\begin{equation}
   \varepsilon = 0.05; \qquad d = \sqrt{3} \pi /2 = 2.721\ldots; \qquad
   k_z = (1 -  j^2/3\pi^2)^{1/2} =  0.772\ldots,
   \label{eq:pars}
\end{equation}
and $j = 4.201\ldots,$ the first positive zero of $J_3'.$ 
The reason for the particular choices (\ref{eq:pars}) is explained later.
The first two terms of (\ref{eq:braid}) are order +3 Bessel beams of the form (\ref{eq:bessel}), the first propagating in $+z$ and centered on the line $x = -d, y = 0;$ the second propagates in $-z,$ and is centered on $x = +d, y = 0;$ the order 3 topological current in both beams is in the $+z$ direction.
The third summand is a plane wave propagating in the $-y$-direction. Since these three waves propagate in very different directions ($+z, -z$ and $-y$), the braid function (\ref{eq:braid}) is far from the paraxial regime.

\begin{figure}
\begin{center}
\includegraphics*[width=8cm]{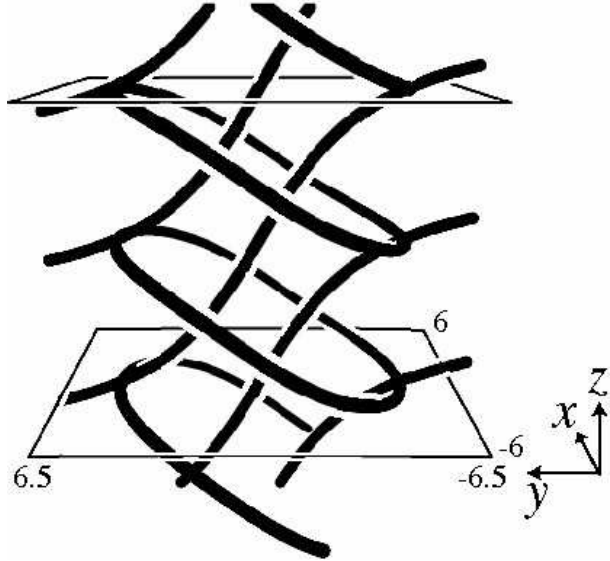}
\end{center}
    \caption{Configuration of the unperturbed braid field $\psi_{\text{un}}$ (\ref{eq:unbraid})  (that is, the braid field $\psi_{\text{braid}}$ (\ref{eq:braid}) with the plane wave amplitude $\varepsilon = 0$). The two horizontal planes are at $z=0$ and $2\pi/k_z$ (one $z$-period). The viewpoint is from the $-x$-direction.}
    \label{fig:unp}
\end{figure}

\begin{figure}
\begin{center}
\includegraphics*[width=8cm]{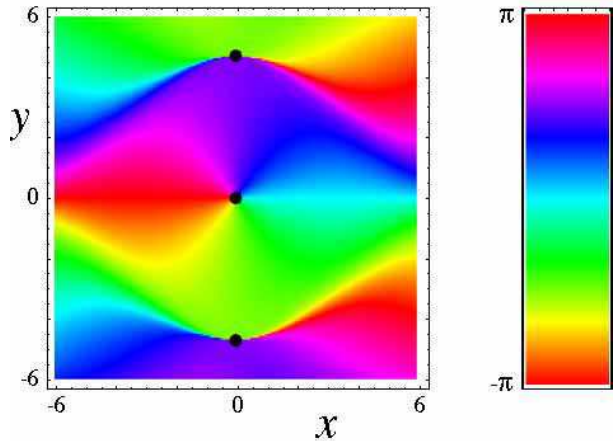}
\end{center}
    \caption{Phase pattern of the unperturbed configuration (\ref{eq:unbraid}) in the plane $z=0.$ There are phase singularities at $(x,y) = (0,0)$ and $(0,\pm 3\pi/2).$}
    \label{fig:unpz0}
\end{figure}

I begin by describing the nodal configuration of $\psi_{\text{braid}}$ without the perturbing plane wave $\mathrm{i} \varepsilon \exp(-\mathrm{i} y),$ that is,
\begin{equation}
   \psi_{\text{un}}(x,y,z) =
   \psi_3(k_z, x + d, y, z) + \psi_3(-k_z, x - d, y, z).
   \label{eq:unbraid}
\end{equation}
The configuration of nodal lines of (\ref{eq:unbraid}) in the vicinity of the $z$-axis is shown in figure \ref{fig:unp}. It shows a series of tilted singularity rings encircling lines confined to the $x = 0$ plane. The rings meet the threading lines at nongeneric connections, at
\begin{equation}
   x = 0, \qquad y = \pm \sqrt{3} d = \pm 3\pi/2; \qquad
   z = 0, \pi/k_z \mod 2\pi/k_z.
   \label{eq:crossings}
\end{equation}
The phase structure in the $z=0$ plane is shown in figure \ref{fig:unpz0}. 
There is one singularity at the origin (this is one of the threading curves in figure \ref{fig:unp}), and two others at $x=0, y = \pm 3\pi/2.$ 
These latter nodes are nongeneric, since not only are they zeros of the field, but also places where the $x$-derivative vanishes. 
This reflects the fact that in three dimensions, they are crossings (\ref{eq:crossings}), shown in figure \ref{fig:unp}. 
The position of phases around the nongeneric singularities is the same for each (e.g. $\mp \pi/2$ in $\pm y$ directions, phase increases in the same sense around each). 

The configuration of singularities of figure \ref{fig:unp} appears characteristic of the nodal pattern of superposed non-coaxial Bessel beams which are counterrotating (i.e. counterpropagating with parallel topological currents), and contains the basic structure of the three strand figure-8 $xy$-trajectory desired for the braid. 
However, it is only with the particular choices of (\ref{eq:pars}) that the crossings occur at precisely the same $z$ value, which is convenient for the present purpose; increasing $d$ slightly makes the two crossings migrate away from the $z=0$ plane (and each other). 
$d$ and $k_z$ were chosen in (\ref{eq:pars}) so the crossings occur in the same planes $z = 0, \pi/k_z$ (mod $2\pi/k_z$), and with a spacing of $3\pi$ in the $y$-coordinate. 
The crossings (\ref{eq:crossings}) are broken when the perturbing plane wave $\mathrm{i} \varepsilon \exp(-\mathrm{i} y)$ is included in (\ref{eq:braid}). 

\begin{figure}
\begin{center}
\includegraphics*[width=8cm]{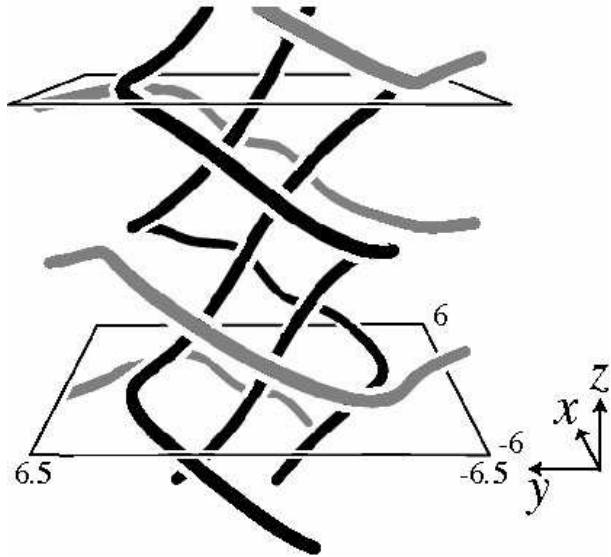}
\end{center}
    \caption{Perturbed braid configuration, where the field of figure \ref{fig:unp} has been perturbed by the plane wave $0.05 \mathrm{i} \exp(-\mathrm{i} y).$ The black curves form the braid plotted in figure \ref{fig:braid}, the grey curves represent other nodal lines. As with figure \ref{fig:unp}, the distance between the two horizontal planes is one $z$-period $2\pi/k_z.$}
    \label{fig:pert}
\end{figure}

\begin{figure}
\begin{center}
\includegraphics*[width=8cm]{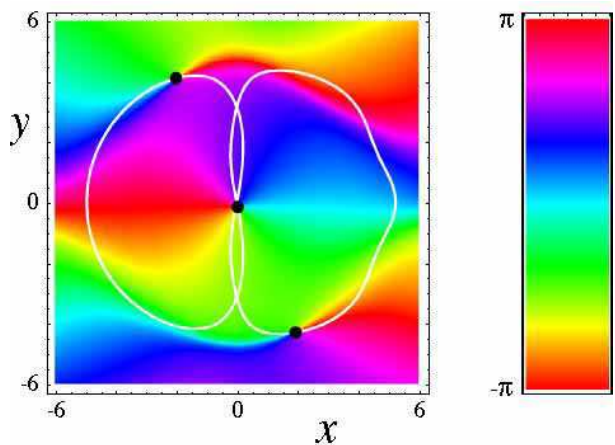}
\end{center}
    \caption{Phase pattern of the full braid field $\psi_{\text{braid}}$ in the plane $z=0.$ The figure-8 $x,y$ trajectory of the braiding singularities is included. The singularities depicted as grey in figure \ref{fig:pert} do not cross the $z=0$ plane, and so are not seen here.}
    \label{fig:pertz0}
\end{figure}

The pair of crossings at $z = 0$ in (\ref{eq:crossings}) are separated by a distance of $3\pi;$ that is, one and a half wavelengths of the plane wave. 
Thus, the phases of the plane wave at the positions of the two $z=0$ crossings of $\psi_{\text{un}}$ differ by $\pi.$ 
This causes the two crossings to break in opposite senses (figure \ref{fig:pert}), and the two zeros in the $z=0$ plane of the braid field are perturbed in opposite directions, as in figure \ref{fig:pertz0}.
Moreover, at $z = \pi/k_z$ (half the $z$-period), all of the phases of $\psi_{\text{un}}$ are reversed.
The plane wave, however, is not changed since its propagation is in the $y$-direction alone; this implies the sense in which the crossing is broken by the plane wave is opposite to that of the corresponding crossing at $z = 0;$ the sense of crossing breakings alternates as $z$ increases.
The full set of nodal lines near the $z$-axis, shown in figure \ref{fig:pert}, includes the braid, shown in black;  there are other nodal lines, shown in grey, which have broken away from the configuration of figure \ref{fig:unp}, and do not take part in the braiding.

The braid described here is structurally stable; if the numerical values of the parameters in (\ref{eq:pars}) are changed slightly, the braid remains topologically unchanged, although past certain critical values of $k_z, d,$ and $\varepsilon,$ it will be dissolved through various reconnection events, as with nodal knots \cite{bd:333}. 
Structural stability means that parameters such as $d$ and $k_z$ need not be precisely controlled in an experimental or numerical realisation.

The key feature of the braid realisation is that the trajectory of the strands of a borromean braid is topologically a figure-8, which has been achieved by wave dislocations in superposed off-axis counterrotating beams. 
The parameters $d$ and $k_z$ were chosen in (\ref{eq:pars}) so that the perturbing plane wave would have opposite phases at the two reconnection points with $z = 0;$ the reconnections were separated by $3\pi.$ 
However, this argument works for any separation $\pi (n + 1/2),$ for $n = 1, 2, \ldots,$ for which the values of $d$ and $k_z$ in (\ref{eq:pars}) should be replaced by
\begin{equation}
   d = \pi (n + 1/2)/\sqrt{3}; \quad k_z = (1- 3j^2/\pi^2 (2n+1)^2)^{1/2}.
   \label{eq:gen}
\end{equation}
In (\ref{eq:pars}), $n = 1$ (if $n = 0,$ then $k_z^2 < 0,$ and the wave is not periodic in $z$). 
In this case, $\varepsilon$ can be arbitrary but small; the braid structure is more sensitive to the size of perturbation for higher $n.$
$\psi_{\text{braid}}$ can straightforwardly be generalised to give other $m$-stranded braids whose transverse trajectory is a figure-8, by using Bessel beams of odd order $m > 3,$ as follows.
In the unperturbed situation, with 3 replaced by $m,$ there is a series of $m$ zero points in the $z = 0$ plane on the $x = 0$ line; $k_z$ and $d$ are chosen such that the two furthest from the origin are at $y = \pm \pi (n+1/2),$ and that the $x$-derivative vanishes there also.
If $m$ is even, the extremal $y$ zeros must be separated by an integer number of wavelengths, and the perturbed pattern, instead of giving a braid, gives a pair of $m/2$-stranded helices.
The construction cannot be generalized in any obvious way to give other, more general braids.

Bessel beams and plane waves are often used in optics, but it is not entirely clear what sort of quantum wave would correspond to (\ref{eq:braid}). 
Would it therefore be possible to create a close enough approximation to $\psi_{\text{braid}}$ in an optical experiment?
This leads to the following observations.
Electromagnetic waves are of course vector fields, and, in general, phase singularities do not occur in vector waves, especially in the nonparaxial regime \cite{nh:wavestructure}; nodes generically only occur in scalar components of the wave. 
In the present case, if all three wave beams in (\ref{eq:braid}) are linearly polarized in the 
$x$-direction, which is possible since the separate wave beams are propagating in the $\pm z$- and $-y$-directions only, the braided nodal lines appear in the $x$-component of the total field (there will also be a longitudinal $z$-component arising from the Bessel beams).
A more serious problem in optical realization is the departure from paraxiality; not only in the interference of coherent counterpropagating beams, but in the fact that in the Bessel beams $k_z \approx 0.8,$ meaning that the divergence of the beam is large. 
Moreover, the braid period is on the scale of the beam period itself, which is extremely small at optical frequencies - unlike the knotted singularity construction of Ref.\ \cite{bd:332}, the absolute scale of the braid structure is fixed.
However, it is possible that use of a higher $n$ as in (\ref{eq:gen}) may lead to a more experimentally realistic beam.
It may be possible to use fields in a different part of the electromagnetic spectrum, such as radio waves, for which Bessel beams can be synthesized and studied \cite{smn+:holograms}. 

The present work shows how potentially realizable beams may be superposed to form a field containing structurally stable, braided nodal lines. 
However, this realization of a braid is not necessarily the only possible one in interfering waves. 
For instance, it would be interesting to investigate whether periodic braiding structure is possible in the space above a two dimensional grating. 
Such a structure would have the desired $z$-periodicity, although it is not composed of  counterpropagating waves as used here. 
Braids, like knots, are particular cases of a wide range of topologically interesting phenomena that may occur in three dimensional wave fields, and one may ask whether other nodal configurations, such as chains, are possible in beam superpositions.
The existence of braided nodal lines widens the topological possibilities of singular optics and quantum vortices, and provides a technical challenge for experiment.

I am grateful to John Hannay and Michael Berry for discussions. 
The figures were generated using {\itshape Mathematica}; figure \ref{fig:braid} with Mitchell Berger's \texttt{tuba} package.
This work was supported by the Leverhulme Trust.


\end{document}